\documentclass[preprint,12pt]{elsarticle}
\usepackage{amssymb}
\usepackage{graphicx}

\journal{}

\begin{document}

\title{ Prediction of half metallic properties in $Ti_{2}CoSi$ Heusler alloy based on density functional theory.}
\author[label1]{A. Birsan}
\author[label1]{P. Palade}
\author[label1]{V. Kuncser}
\address[label1]{National Institute of Materials Physics, PO Box MG-07, Bucharest, Romania}

\begin{abstract}
The electronic and magnetic properties of $Ti_{2}CoSi$ Heusler compound are investigated using density functional calculations. The optimized lattice constant  is found to be 6.030$\dot{A}$. The compound is a half-metallic ferromagnet with an energy gap in minority spin channel of 0.652 eV  at equilibrium lattice constant, which leads to a $100\%$ spin-polarization. The obtained total magnetic moment from spin-polarized calculations is 3.0 $\mu_{B}$ for values of lattice constants higher than  5.941 $\dot{A}$. The half-metallicity is spoiled for a compressed volume of $4\%$, suggesting a possible application as pressure sensitive material.
\end{abstract}


\maketitle

\section{Introduction}
\label{Introduction}
Heusler compounds are ternary intermetallics, known since 1903 \citep{Heusler1903}, when was proved that $Cu_{2}MnAl$ compound has magnetic properties, even though by itself, none of its constituents is ferromagnetic. 

From the structural of view, the large Heusler family are described by only two variants: the so-called full-Heusler $X_{2}YZ$ phases which typically crystallize in $Cu_{2}MnAl$ ($L2_{1}$) -type structure and the half-Heusler $XYZ$ variants with $C1_{b}$ structure;  $X$ is a transition metal, $Y$ may be a transition metal or a rare-metal and $Z$ is a main group element. In full Heusler alloys, if the atomic number of $Y$ is higher than the one of $X$, making the $Y$ element more electronegative than $X$, is observed  the inverse Heusler structure of $Hg_{2}CuTi$-type with $F\bar{4}3m$ space group and $X$ atoms placed on the Wyckoff positions 4a(0,0,0) and 4c(1/4,1/4,1/4), $Y$ and $Z$ in 4b(1/2,1/2,1/2) and 4d(3/4,3/4,3/4) respectively \citep{Kandpal2007}.

The properties of Heusler compounds can be altered by substitution of elements, however no single set of properties can characterize the entire Heusler family. Due to all kind of magnetic behavior and multifunctional properties, Heusler alloys are expected to play an important role  in the research field of spintronics \citep{Felser2007}, magneto-optical reading and recording devices \citep{vanEngen1983}, magnetic tunnel junctions or tunneling magnetoresistance devices \citep{Moodera1999}. 

The half-metallic properties, discovered first by de Groot and his collaborators in $NiMnSb$ \citep{Groot1983} characterize the large family of magneto-electrical Heusler compounds. Half-metallic materials being hybrids between metals and semiconductors or isolators, can provide a fully spinpolarised current due to metallic behavior for electrons of one spin orientation and the semiconducting nature for electrons with opposite spin orientation \citep{Boeck2002}. 
Numerous studied of structure-to-properties relations of $Mn_{2}$ \citep{Wei2011}, $Fe_{2}$ \citep{Luo2012,Li2012,Kervan2012INTERM} and $Co_{2}$ \citep{Wurmehl2006PRB,Kandpal2007JMMM,Marukame2006,Tezuka2006,Galanakis2002}-based Heusler compounds provide an insight into this outstanding class of materials. The $Co_{2}$-based Heusler compounds, due to their Curie temperatures are used today in magnetic tunnel junctions \citep{Wurmehl2006,Wang2010,Sargolzaei2006} while $Mn_{2}$ -based compounds might be incorporated in spin torque devices because of perpendicular magnetic anisotropy, achieved in thin films \citep{Wu2009}.

Recently, scientific interest was stimulated by theoretical investigations of many $Ti_{2}$ based compounds, $Ti_{2}YZ$ ($Y$=V, Cr, Mn, Fe, Co, Ni, Cu, Zn and $Z$=B, Al, Ga, In, Si, Ge ) \citep{Lei2011,KervanPCS2011,KervanSSC151,Bayar2011,KervanJMMM2012,Huang2012,KervanElMat2012,Zheng2012,Wei2012}, which exhibit half metallic character. However, the electronic configuration of $Ti_{2}CoSi$ compound has never been analyzed. In this article, the comprehensive theoretical study, by means of the self-consistent full potential linearized augmented plane wave (FPLAPW) method is focused on the spin-polarization electronic structure and magnetic properties of the new full Heusler compound, $Ti_{2}CoSi$. Based on the analysis of these properties is predicted that $Ti_{2}CoSi$ system presents half metallic character being suitable in spintronics and pressure sensitive devices.

\section{Method of calculation}
\label{calculation}
Self-consistent electronic-structure calculations were performed using the full-potential linearized augmented plane wave (FPLAPW) method, implemented in WIEN2K \citep{Blaha}. The Perdew Burke Ernzerhof \citep{Perdew} generalized gradient approximation GGA was employed for the exchange and correlation potential; the muffin-tin model for the crystal assumes that the unit cell is divided in two regions, within and outside the muffin-tin spheres. The plane wave cut-off value $K_{max}\*R_{MT} = 7$ was used (where $K_{max}$ is the maximum modulus for the reciprocal lattice vector). The muffin-tin radius $R_{MT}$ were 2.21 atomic units (a.u.) for both Ti and Co and 2.35 a.u. for Si. The energy eigenvalues and eigenvectors of a grid containing 2925k points was considered for integration over the irreducible part of the Brillouin zone (BZ), using the modified tetrahedron method \citep{Blochl}.  The self-consistency was attained when the integrated charge difference between two successive iterations was less than 0.0001e/a.u.3 and the total energy deviation better than 0.01mRy per cell.

\section{Results and Discussions}
\label{results}
\begin{figure}
 \begin{center}
     \includegraphics[scale=0.7]{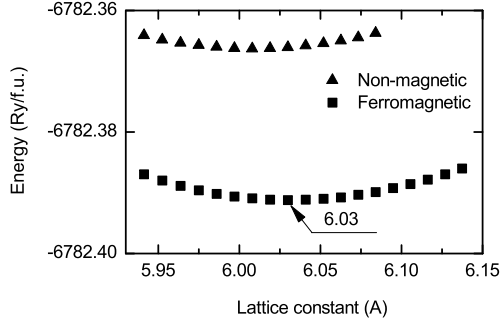}
  \end{center}
    \caption{Structural optimization for $Ti_{2}CoSi$. - the calculated total energy vs. the lattice constant.  }
        \label{fig:optimizareTi2CoSi}  
\end{figure}

To explain the properties of $Ti_{2}CoSi$, the total energy minimization needs to be considered, for estimation of equilibrium lattice constant. In order to define the unit cell of  $Ti_{2}CoSi$ system, for both magnetic and non-magnetic configurations, the total energy is calculated as function of lattice constant for  $Hg_{2}CuTi$ structure. This prototype structure is expected to occur because number of 3d electrons of $Co$ atom exceed that of $Ti_{2}$ atom the $Ti_{2}CoSi$ compound \citep{Kandpal2007}. The ferromagnetic phase has lower energy than the non-magnetic phase as shown in Figure \ref{fig:optimizareTi2CoSi}. The following calculations are all based on the estimated equilibrium lattice constant, which is found to be 6.030 $\dot{A}$ for ferromagnetic phase. 
   
\begin{figure}
 \begin{center}
         \includegraphics[scale=1]{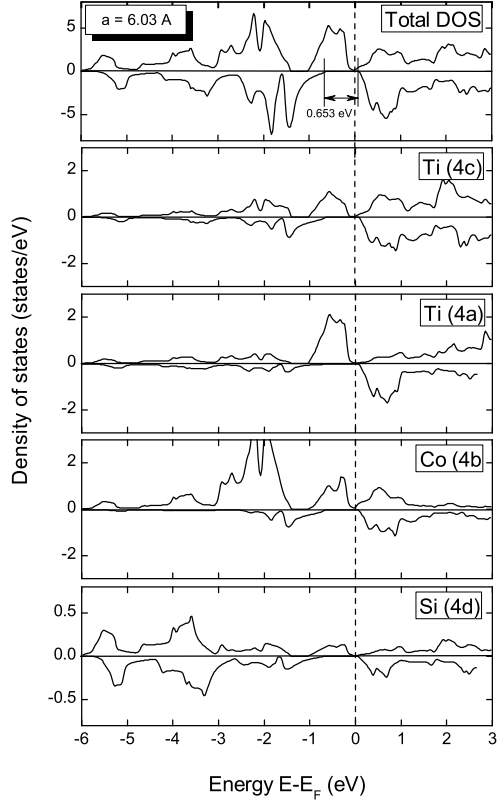}  
    \end{center}
    \caption{The spin projected total DOS and partial DOS calculated at predicted equilibrium lattice constant.}
       \label{fig:totaldosTi2CoSi}
\end{figure}

The spin-polarized calculations for $Ti_{2}CoSi$ compound performed using the equilibrium lattice constant are displayed in Figure \ref{fig:totaldosTi2CoSi}   and 4c, 4a, 4b and 4d represent the Wyckoff positions. 

The compound exhibits  half-metallic nature with a prominent energy gap at the Fermi level in minority spin channel (spin-down band) and metallic character in majority spin channel (spin-up band). The significant contribution to the energy between -3eV and -1eV comes from d electrons of Co atoms, while between -6 eV and -3 eV from p electrons of Si.

\begin{figure}
 \begin{center}
   \includegraphics[scale=1]{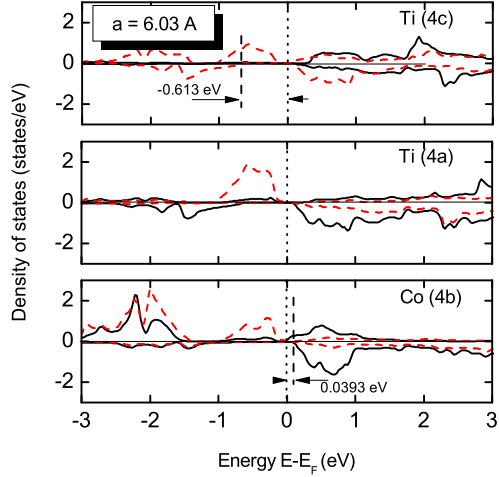}  
 \end{center}
    \caption{The main partial densities of states at optimized lattice parameter of $Ti_{2}CoSi$, Fermi level, $d_{eg}$ and $d_{t2g}$ being indicated by dotted, solid and dashed line, respectively}
    \label{fig:sitedosTi2CoSi}
\end{figure}

The Fermi level for equilibrium lattice parameter is found nearly on the edge of the conduction band. The highest occupied states of valence band are located at -0.613 eV below the $E_{F}$ and belong to triple degenerated states Ti (4c) $d_{t2g}$. The lowest unoccupied states of the conduction band, with a strong Co character, the double degenerated states $d_{eg}$ are situated at 0.0393 eV above the $E_{F}$ (Figure \ref{fig:sitedosTi2CoSi}). As result, the energy gap from minority spin channel is formed due to Ti(4c) - Co hybridization. Besides the Ti(4c)-Co bounding interactions, Ti(4c)-Ti(4a) couplings provide also a strong hybridization around the Fermi level. However, because of the dissimilar next-nearest neighborhood, Ti(4c)-Ti(4a) is different than the typical $X$-$X$ ($X_{2}YZ$) couplings, which determine the band gap in full-Heusler compounds, with $L2_{1}$ structure \citep{Galanakis2002}. 

\begin{figure}
 \begin{center}
    \includegraphics[scale=1]{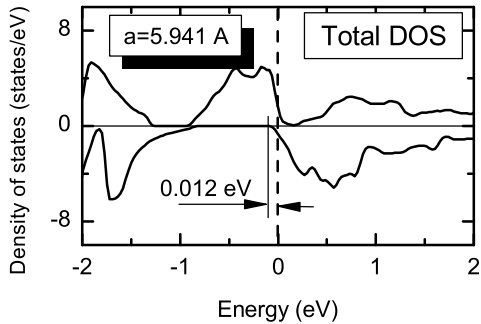}
      \end{center}
   \caption{Electronic structure of $Ti_{2}CoSi$ at 4$\%$ compression of optimized volume.}
    \label{fig:4compressedvol}
\end{figure}

The critical transition point between the metallic and half metallic behavior of the system is found for a 4$\%$ compression of optimized volume which corresponds to the lattice parameter value of 5.941$\dot{A}$. Figure \ref{fig:4compressedvol} shows the total densities of states for the critical transition point, when $E_{F}$ is located  at 0.012 eV distance from the lowest unoccupied states of conduction band. 
 Therefore, if the lattice constant is reduced to a value lower than the lattice constant corresponding to critical transition point the Fermi level doesn't lie into the energy gap from minority spin channel and the $Ti_{2}CoSi$ compound loses its half-metallic nature.
 
\begin{figure}
 \begin{center}
    \includegraphics[scale=0.8]{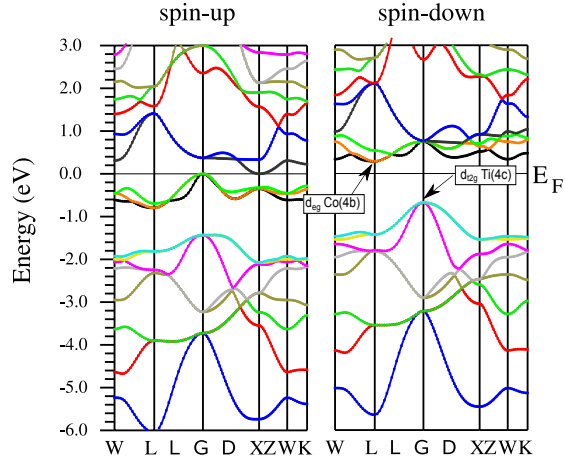} 
      \end{center}
   \caption{The band structure of $Ti_{2}CoSi$ for majority spin channel (spin-up) in the left panel and minority spin channel (spin-down) in the right panel for geometrical optimized structure}
    \label{fig:bandTi2CoSi}
\end{figure}

\begin{figure}
 \begin{center}
    \includegraphics[scale=0.8]{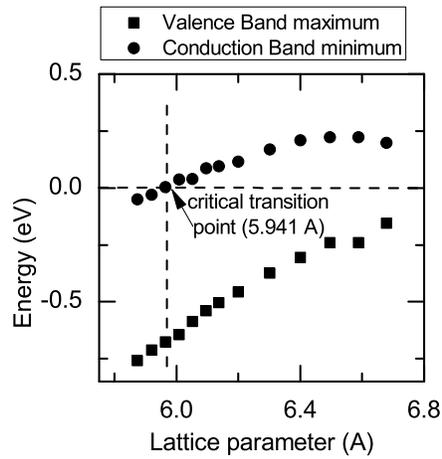}
      \end{center}
   \caption{The positions of highest occupied states from valence band (solid squares) and of lowest unoccupied states from the conduction band (solid circles) of total DOSs (minority spin channel) for $Ti_{2}CoSi$ as function of lattice parameter.}
    \label{fig:gapTi2CoSi}
\end{figure}

The band structure of  $Ti_{2}CoSi$ compound at equilibrium geometry is shown in Figure \ref{fig:bandTi2CoSi}. In the left panel of the figure is plotted the metallic intersection of Fermi level from majority spin channel while in the right panel, the energy band gap from the minority spin channel.  
The band gap is an indirect one with a width of 0.652 eV, calculated between the energy from the highest occupied band at the G point (0.613 eV) and the lowest unoccupied band, at the L point (0.039eV), in the minority spin channel. In $Ti_{2}CoB$ system the width of the band gap predicted is 0.64 eV \citep{KervanSSC151}, in $Ti_{2}CoAl$, 0.49eV \citep{Bayar2011}, $Ti_{2}CoGa$  alloy exhibits a gap of 0.5 eV \citep{KervanJMMM2012}, while in the $Ti_{2}CoGe$ compound the energy gap is 0.6 eV.  \citep{Huang2012} or 0.61 eV \citep{KervanElMat2012}. It is larger than the energy gap reported not only for other $Ti_{2}Co$ based compounds but also $Ti_{2}Ni$ \citep{Lei2011,Zheng2012,Wei2012} and $Ti_{2}Fe$ \citep{KervanPCS2011,Zheng2012,Wei2012} -based Heusler alloys.  
The dependence of the half-metallic behavior on the lattice parameter is plotted in Figure \ref{fig:gapTi2CoSi}. The $Ti_{2}CoSi$ compound presents 100 $\%$ spin polarization above the critical transition point and this starts decreasing while the half metallicity is spoiled. 

\begin{figure}
 \begin{center}
    \includegraphics[scale=1]{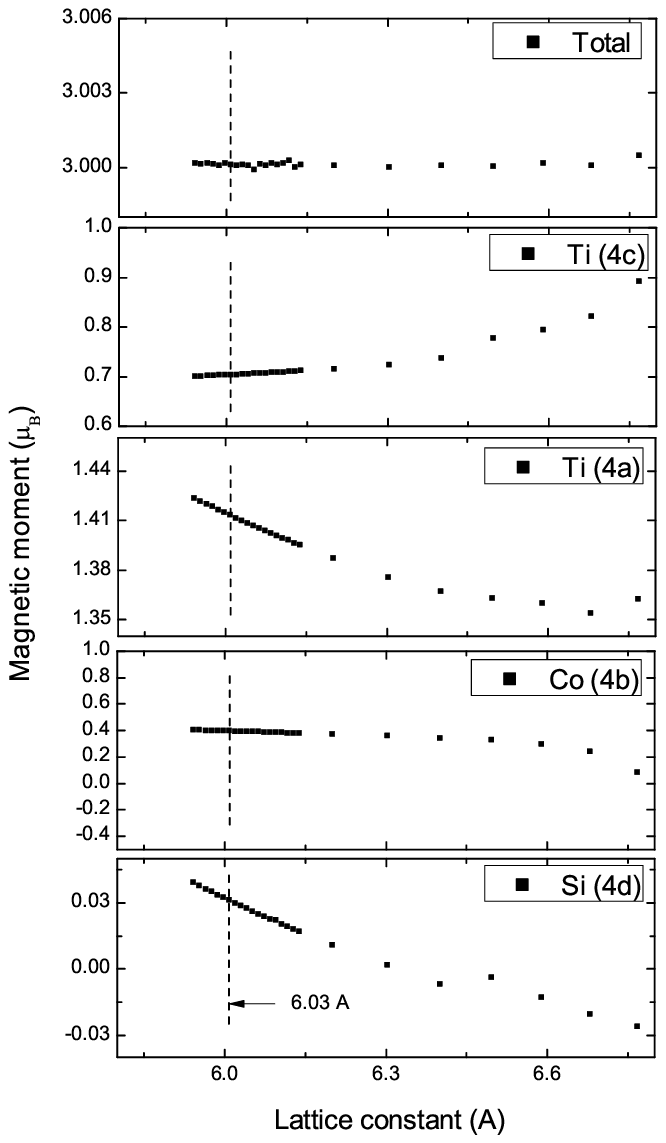}
      \end{center}
   \caption{Total and site-specific magnetic moments of $Ti_{2}CoSi$ vs lattice parameter.}
    \label{fig:magneticmomentTi2CoSi}
\end{figure}

The calculated values of the total spin magnetic moment $M^{calc}_{tot}$ ($\mu_{B}/f.u.$) and the spin magnetic moment of each atom $M^{calc}$ ($\mu_{B}/atom$) as function of lattice parameters are plotted in Figure \ref{fig:magneticmomentTi2CoSi}.
 The total magnetic moment is constant at 3 $\mu_{B}$ for  lattice constant higher than 5.941 $\dot{A}$.  The site-resolved magnetic moments per atom resulted at the optimized lattice constant are 0.705, 1.410, 0.395 and  -0.028 $\mu_{B}$ for Ti(4c), Ti(4a), Co and Si, respectively.  Surprisingly, the highest spin magnetic moments come from Ti atoms which natively don't have magnetic properties. In addition, the influence of different neighborhood determines Ti atoms to have dissimilar magnetic moments: Ti atoms located in 4a - Wyckoff position and surrounded by Si atoms at a distance of 2.74$\dot{A}$, present higher magnetic moments than those with Co atoms neighbors . Similar results were obtained for other $Ti_{2}Co$-based Heusler alloys \citep{KervanSSC151,Bayar2011,KervanJMMM2012,Huang2012,KervanElMat2012}. The atom resolved magnetic moments of Ti (4a), Co and Si decrease with increasing of  lattice constant, while for Ti (4c), it increases. 

\section{Conclusions}
\label{conclusions}
In summary, the density of states, energy bands, and magnetic moments of the new full-Heusler compound, $Ti_{2}CoSi$, based on the inverse Heusler structure  are studied within the framework of the density of states theory with the Perdew – Burke – Ernzerhof generalized gradient approximation (GGA). 
 The energy gap present in the minority spin channel is 0.652 eV at the optimized lattice constant of 6.030$\dot{A}$. The total magnetic moment resulted from spin polarized calculations is 3 $\mu_{B}$ and a fully polarized character, for values of lattice constant higher than 5.941 $\dot{A}$. The main contribution to the magnetic moments comes from Ti atoms. From applications point of view $Ti_{2}CoSi$ compound is predicted to be suitable not only for spintronic devices due to high spin polarization and the small energy separation between the conduction band minimum and the Fermi level, but also as pressure sensitive material, because of transition from half-metal to metal obtained at 4$\%$ compression of optimized volume. 

\section{Acknowledgments}
\label{acknowledgments}
This work was financially supported from the projects  PNII IDEI 75/2011 and Core Program PN09-450103 of the Romanian Ministry of Education Research, Youth and Sport. 
\bibliographystyle{elsarticle-harv}

\end{document}